# Optical signature of symmetry variations and spin-valley coupling in atomically thin tungsten dichalcogenides

Hualing Zeng[1], Gui-Bin Liu[1,4], Junfeng Dai[2,1], Yajun Yan[3], Bairen Zhu[1], Ruicong He[1], Lu Xie[1], Shijie Xu[1], Xianhui Chen[3], Wang Yao[1,4+], Xiaodong Cui[1*]

1. Department of Physics, The University of Hong Kong, Pokfulam road, Hong Kong, China

2. Department of Physics, South University of Science and Technology of China, Shenzhen, Guangdong, China

3. Hefei National Laboratory for physical Science at Microscale and Department of Physics, University of Science and Technology of China, Hefei, Anhui 230026, China.

4. Center for Theoretical and Computational Physics, The University of Hong Kong, Hong Kong, China

[+]Email:   wangyao@hku.hk

[*]Email:   xdcui@hku.hk

**Motivated by the triumph and limitation of graphene for electronic applications[1], atomically thin layers of group VI transition metal dichalcogenides are attracting extensive interest as a class of graphene-like semiconductors[2,3] with a desired band-gap in the visible frequency range[4,5]. The monolayers feature a valence band spin splitting with opposite sign in the two valleys located at corners of 1st**

**Brillouin zone[6]. This spin-valley coupling, particularly pronounced in tungsten dichalcogenides, can benefit potential spintronics and valleytronics with the important consequences of spin-valley interplay and the suppression of spin and valley relaxations[6]. Here we report the first optical studies of $WS_2$ and $WSe_2$ monolayers and multilayers. The efficiency of second harmonic generation shows a dramatic even-odd oscillation with the number of layers, consistent with the presence (absence) of inversion symmetry in even-layer (odd-layer). Photoluminescence (PL) measurements show the crossover from an indirect band gap semiconductor at mutilayers to a direct-gap one at monolayers. The PL spectra and first-principle calculations consistently reveal a spin-valley coupling of 0.4 eV which suppresses interlayer hopping and manifests as a thickness independent splitting pattern at valence band edge near K points. This giant spin-valley coupling, together with the valley dependent physical properties [6-11], may lead to rich possibilities for manipulating spin and valley degrees of freedom in these atomically thin 2D materials.**

The family of group VI transition metal dichalcogenides $MX_2$ (M=Mo,W; X=S,Se) has a structure of X-M-X covalently bonded hexagonal quisi-2D network staked by weak Van der Waals forces. Materials in this family have similar band structures as well as electric and optical properties. *ab initio* calculations predict that $MX_2$ exhibits a transition from an indirect-gapped semiconductor in multilayer form to a direct band-gap one at visible range in monolayer, which has been experimentally verified in MoS$_2$[4,5]. MX$_2$ thin films exhibit 2H stacking order: the neighboring layers are 180 degree in plane rotation of each other with the metal atom of a given layer sits exactly on top of the chalcogenide atom of adjacent layer (see Figure 1a). There is an even-odd variation in the structural symmetry of ultrathin films: inversion symmetry is absent (present) in films with odd (even) number of layers with space group of $D^1_{3h}$ ($D^4_{6h}$).

MX$_2$ monolayer, the elementary unit to form ultrathin films by weak stacking, features a novel spin-valley coupled band structure[6]. At the corners of the 1$^{st}$ Brillouin zone, the valence (conduction) band has two inequivalent valleys described by massive Dirac fermions. Owning to the broken inversion symmetry in monolayers, the strong spin-orbit coupling from the *d*-orbitals of metal atom results in a valence band spin splitting at K points, with a magnitude as large as ~ 0.4 eV in tungsten dichalcogenides[6,12]. The spin-splitting has opposite signs at the K and K' valleys as they are time reversal of each other. This *spin-valley coupling* forms the basis for manipulation of spin and valley degrees of freedom in these novel 2D semiconductors when combined with valley contrasted electric, magnetic and optical properties arising from inversion symmetry breaking[6-11].

Here we report our experimental study on optical properties of ultrathin WS$_2$ and WSe$_2$ mono-, bi-, tri- and quad-layer samples by means of Raman scattering, second harmonic generation (SHG) and photoluminescence (PL). The efficiency of SHG at normal incidence on WS$_2$ and WSe$_2$ monolayers and multilayers shows a dramatic even-odd oscillation with the number of layers: negligible at even-layer and nonzero at odd-layer, with maximum strength at monolayers. PL measurements demonstrate that $WS_2$ and $WSe_2$ exhibit a transition from an indirect-gap semiconductor at multi-layers to a direct-gap one at monolayers with an enhancement of the PL quantum efficiency (QE) at a factor of more than $10^3$ compared to bulk samples. Remarkably, a weak emission peak (B) is observed at an energy ~ 0.4 eV higher than the prominent direct bandgap transition peak (A) in all monolayer and multilayer samples. For both $WS_2$ and $WSe_2$, the magnitude of A-B splitting is independent of the number of layers and coincides with the spin-valley coupling strength in monolayers. *ab initio* calculations show that this thickness independent splitting pattern is a direct consequence of the giant spin-valley coupling which fully suppresses interlayer hopping at valence band edge at K points because of the sign change of the spin-valley coupling from layer to layer in the 2H stacking order.

$WS_2$ and $WSe_2$ flakes were mechanically exfoliated from synthesized single crystal bulk samples onto silicon wafers capped with a 300nm thick $SiO_2$ by a method analogous to the way of producing graphene[1]. $WS_2$ and $WSe_2$ slabs were first visually screened with interference color through optical microscope. Typical optical images of $WS_2$ and $WSe_2$ ultrathin slabs are presented in Figure 2.a and 2.e. The film thickness is confirmed by atomic force microscope. PL spectra are also used as an indicator of monolayer samples (Figure 2.b and Figure 3.a; Figure 2.f and Figure 3.d).

Raman scattering was carried out with a confocal setup. For $MX_2$ layered compounds, there are generally four Raman-active modes, namely $A_{1g}$, $E_{1g}$, $E_{2g}^1$ and

$E_{2g}^2$ modes[13,14]. $E_{1g}$ mode and low energy $E_{2g}^2$ mode are absent in our measurements due to the forbidden selection rule in the back-scattering geometry and the limited rejection against Rayleigh scattering respectively. The presented study focuses on the in-plane vibrational $E_{2g}^1$ mode and the out-of-plane vibrational $A_{1g}$ mode. As these two modes are both polarization sensitive, the exciting laser line was tuned to an unpolarized state. Figure 1.b~e. present the representative Raman spectra of $WS_2$ and $WSe_2$ slabs with layer number N=1 to 4 and bulk. In the case of $WS_2$, we observe the $E_{2g}^1$ mode at ~350 cm$^{-1}$ and the $A_{1g}$ mode at ~420 cm$^{-1}$ [15] (Figure 1.b). The $E_{2g}^1$ mode shows little dependence on the film thickness, while the $A_{1g}$ mode undergoes a blue shift with increasing layer number, showing a lattice stiffening effect as expected when additional layers are added. By examining the frequency differences (Δω) between the $E_{2g}^1$ mode and $A_{1g}$ mode, the sample thickness could be identified accordingly. As indicated in Figure 1.e showing the frequency difference as a function of layer number N, we label Δω= 65.5cm$^{-1}$, 68.3cm$^{-1}$ and 69.2cm$^{-1}$ to monolayer, bilayer and trilayer respectively. For slabs composed of four and more layers, Δω converges to the bulk value at around 70cm$^{-1}$. Notably, from monolayer to trilayer the $A_{1g}$ peak is roughly 0.5, 1 and 1.8 times the height of the $E_{2g}^1$ peak (Figure 1.c), demonstrating that the ratio of the intensity of $A_{1g}$ mode to that of $E_{2g}^1$ mode could also be used as an indicator of sample thickness. For $WSe_2$, two dominant peaks are observed in the spectral range of 100cm$^{-1}$ to 300cm$^{-1}$ in various samples from monolayer to bulk (Figure 1.d). However, little systematic trend could be observed on both the two modes as shown in Figure 1.e.

An experimental method to examine the inversion symmetry in ultrathin film is to study the nonlinear optical effect such as SHG determined by the second order susceptibility $\chi^{(2)}$[16]. In the presence of inversion symmetry, $\chi^{(2)}$ is zero. A dramatic even-odd oscillation pattern is indeed observed on the SHG intensity consistent with the presence (absence) of inversion symmetry in even-layer (odd-layer) as shown in Figure 2.c and Figure 2.g. $WS_2$ and $WSe_2$ ultrathin slabs are scanned by a 150fs pulsed laser beam with a wavelength of 800nm at normal incidence and the signal at the double frequency (400nm) is collected. As expected, negligible SHG are observed in both $WS_2$ and $WSe_2$ slabs with even layer number or bulk samples, and strong second harmonic emission arises from multilayer slabs with odd layer number. Notably, the brightest second harmonic emission is observed in monolayers of both WS2 and WSe2. The intensity of the second harmonic emission decays gradually with the increasing layer number, as indicated in Figure 2.d and Figure 2.h. This descending trend can be understood as an averaging effect when additional layers lower the structure asymmetry as a whole.

The Photoluminescence study shows that $WS_2$ and $WSe_2$ exhibits a transition from indirect band-gap semiconductor in the form of bulk and multilayers to direct band-gap one in monolayers, similar to $MoS_2$[4,5,17,18]. Figure 3 illustrates the PL spectra of $WS_2$ and $WSe_2$ samples with various thicknesses measured under the same condition with an excitation at 2.41eV. Insets in Figure 3.a and 3.d show the PL peak intensity as a function of thickness. The PL intensity is found to be extremely weak on bulk samples, consistent with an indirect band-gap semiconductor in bulk form. As WS$_2$ and WSe$_2$ thin to a few atomic layers, the intensity of PL from direct interband transition dramatically increases and reaches maximum at monolayers, more than 3 orders of magnitude stronger than that from bulk. Both WS$_2$ and WSe$_2$ monolayers show much brighter PL with intensity at one order of magnitude higher than bilayers. The peak originating from the indirect band-gap transition (labeled as "I" in Figure 3.b and 3.e) gradually shifts toward higher energy and fades to null at monolayers. These behaviors are fully consistent with the calculated band structures (see Figure 4).

Besides the peaks from indirect transition and the prominent direct transition peak (A), weak PL peak (B) is observed at higher energy in WS$_2$ and WSe$_2$ at all thickness. Strikingly, the splitting between A and B peaks are almost *identical*, around 0.4eV for mono-, bi-, tri- and quad-layer samples (see Figure 3.c and 3.f). In monolayers, it is now well understood that the valance band edges at K points have a spin splitting purely arising from the strong spin-orbit coupling in the *d*-orbitals of the *W* atom, and we can attribute A and B to the direct-gap transitions between the spin split valence bands and the conduction band at the K points. [4] However, in multilayers, both the spin-orbit coupling and the interlayer hopping contribute to the valence band splitting at K points. Besides, even layer samples are inversion symmetric while odd layer samples are asymmetric. These in general would result in complex splitting patterns in multilayers.

To understand the A-B splitting pattern in WS$_2$ and WSe$_2$ ultrathin layers, we perform *ab initio* calculations of the band structures using the projector augmented wave method[19] and generalized gradient approximation[20] implemented in the ABINIT code[21,22]. The structure parameters are taken from Ref[12]. Figure 4.a-d (top row) shows the band structures of mono-, bi-, tri-, and quad-layer WS$_2$ in the *absence* of SOC. As expected, the valance band edge at K point splits into two, three, and four bands respectively for bi-, tri-, and quad-layer WS$_2$ due to the interlayer hopping. A hopping matrix element $t \sim 0.1$ eV can be extracted from the splitting pattern. However, when SOC is included, the splitting pattern is completely changed as shown in Figure 4.e-h. The valence band edges split into two degenerate manifolds with a splitting magnitude independent of the layer thickness for both WS$_2$ and WSe$_2$. The band structures of WSe2 layers could be found in supplementary information. This is in perfect agreement with the A-B splitting patterns observed in the photoluminescence of mono-, bi-, tri-, and quad-layer WS$_2$ and WSe$_2$. The calculated valence band edge splittings of 0.43eV in WS$_2$ and 0.47eV in WSe$_2$ also agree with the measured A-B splitting of 0.4eV as shown in Figure 3 and the supplementary information.

In fact, the unexpected splitting patterns in multilayer WS$_2$ and WSe$_2$ are manifestations of the giant spin-valley coupling in valence band[6]. In monolayers, the Kramer's doublet $|K\uparrow\rangle$ and $|K'\downarrow\rangle$ are separated from the other doublet $|K'\uparrow\rangle$ and $|K\downarrow\rangle$ by the spin-valley coupling energy of $\lambda_{svc} \sim 0.4$ eV.[6] In the 2H stacked multilayers, any two neighboring layers are 180 degree in plane rotation of each other. This rotation switches K and K' valleys but leaves the spin unchanged, which results in a sign change for the spin-valley coupling from layer to layer. Thus, the spin-conserving interlayer hopping can only couple states in neighboring layers with a detuning $\lambda_{svc}$. Interlayer hopping is therefore strongly suppressed by the giant spin-valley coupling. A direct consequence is that the splitting patterns remain the same as that of monolayers, and the valence band Bloch states near K points are largely localized in individual layers, as if the interlayer hopping is absent. This is indeed confirmed by *ab initio* calculations of electron density distributions of these Bloch states (see Figure 5 and table S2). The full suppression of interlayer hopping at K points by the spin-valley coupling is unique to tungsten dichalcogenides where $\lambda_{svc} \gg t$. For molybdenum dichalcogenides where $\lambda_{svc} \sim t$, interlayer hopping will manifest in both the splitting pattern and Bloch function[23].

In Summary, we study the optical properties of ultrathin tungsten dichalcogenide atomic layers by means of Raman scattering, second harmonic generation and photoluminescence. SHG efficiency shows a dramatic even-odd oscillation pattern consistent with the presence (absence) of inversion symmetry in films with even (odd) number of layers. The PL spectra and first-principles calculations consistently reveal a giant spin-valley coupling of ~ 0.4 eV. This giant spin-valley coupling may offer a potential measure to manipulate spin and valley degrees of freedom in tungsten dichalcogenide monolayers.

**Acknowledgement**

We thank Mr. Xiaohu Wang, Dr. Changcheng Zheng and Dr. Helen Leung for technique assistance. The work was supported by University Grant Council (AoE/P-04/08) and the Research Grant Council (HKU706412P) of the government of Hong Kong SAR and the National Basic Research Program of China (973 Program, Grant No. 2012CB922002).

**Author contributions**: X.C. and H.Z. designed the experiments. X.Chen. and Y.Y. grew the single crystal $WS_2$ and $WSe_2$, H.Z., J.D., B.Z., R.H., and L.X. performed the experiments. W.Y. and G.L. contributed to the theoretical interpretations and *ab initio* calculations. All authors discussed the results and co-wrote the paper.

**Figure**

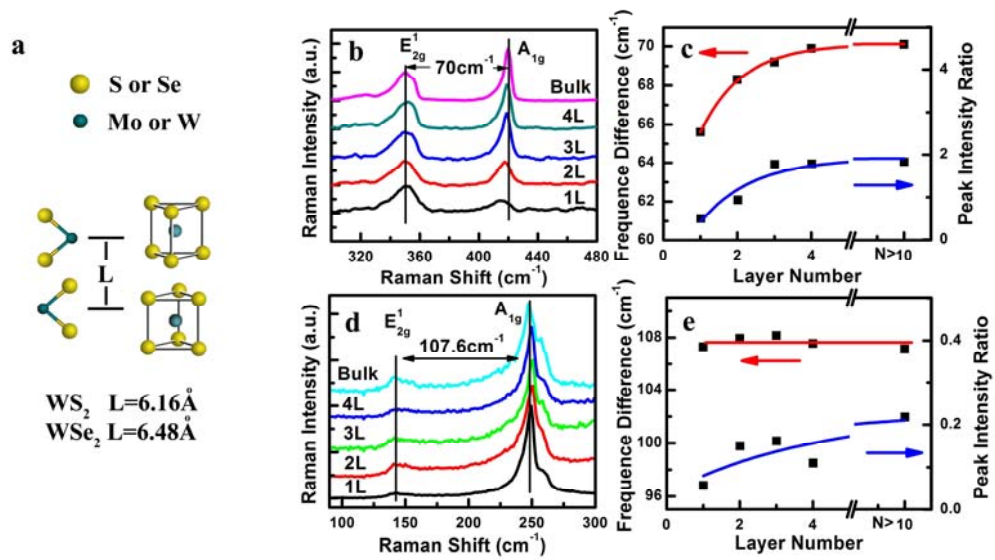

Figure 1. (a) Schematic of bulk MX$_2$ structure and unit cell. (b) and (d): Raman spectra of WS$_2$ (b) and WSe$_2$ (d) ultrathin layers. (c) and (e): The frequency difference (red) and the peak intensity ratio (blue) between $E^1_{2g}$ and A$_{1g}$ modes as a function of layer thickness in WS$_2$ (c) and WSe$_2$ (e) respectively.

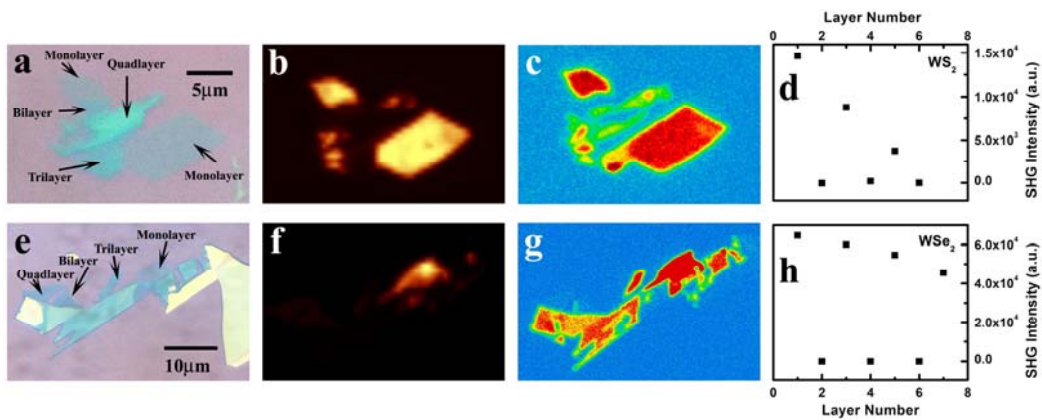

Figure 2. Second harmonic generation on WS$_2$ (top panels) and WSe$_2$ (bottom panels) ultrathin slabs. (a) and (e): optical images of WS$_2$ (a) and WSe$_2$ (e) slabs on Si substrates with 300nm SiO$_2$ cap-layer. (b) and (f): photoluminescence images at direct gap transition energy of the corresponding WS$_2$ (b) and WSe$_2$ (f) slabs excited at 2.41eV. Only monolayers are visible at the present contrast. (c) and (g): the corresponding SHG under a 800nm excitation at normal incidence (150fs, 80MHZ) on WS$_2$ (c) and WSe$_2$ (g) respectively. The highest intensity labeled in red arises from monolayers. (d) and (h): The relative intensity of SHG as a function of the layer

thickness in $WS_2$ (d) and $WSe_2$ (h). The SHG shows an even-odd oscillation dependence on the layer number.

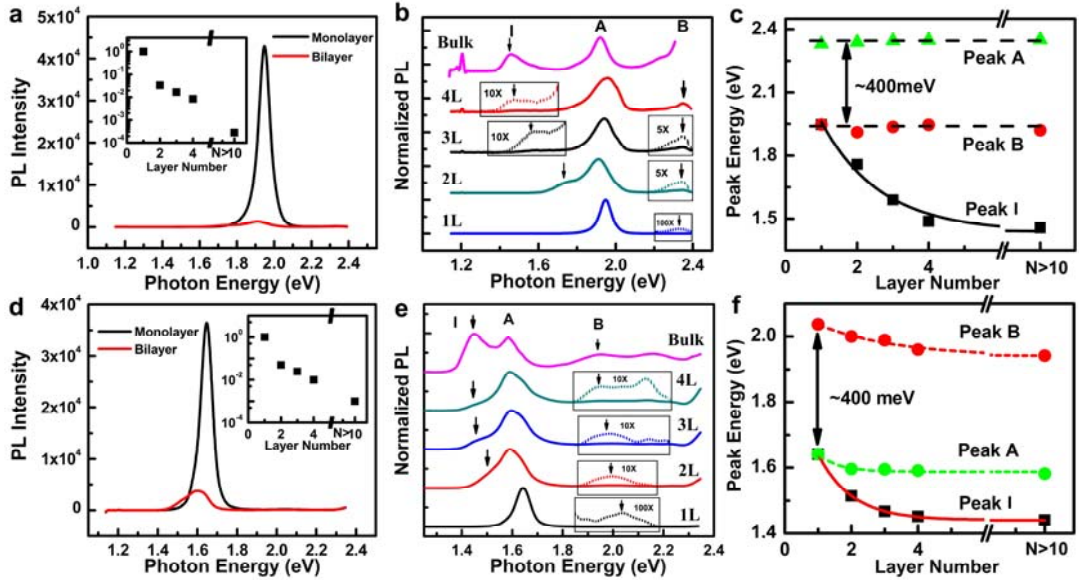

Figure 3. Photoluminescence spectra of $WS_2$ (top panels) and $WSe_2$ (bottom panels) ultrathin layers. (a) and (d): PL spectra from $WS_2$ (a) and $WSe_2$ (d) monolayers and bilayers respectively. The spectra were taken at the same conditions (excitation power, exposure time, *etc*.). The PL intensity of monolayers is one order of magnitude higher than that of bilayers. Insets present the relative PL intensity of $WS_2$ (a) and $WSe_2$ (d) multilayers respectively as a function of layer thickness at the same conditions (set the PL intensity of monolayer at 1). (b) and (e): The normalized PL spectra (with respect to the peak A) of $WS_2$ (b) and $WSe_2$ (e) ultrathin layers. I labels the luminescence from indirect gap interband transition, and A and B label the direct-gap transitions from the split valence band edge to the conduction band edge at K points (see text). Spectra (dash line) in the zoom windows have been multiplied by a factor as indicated for clarity. (c) and (f): The peak positions of I, A and B transitions as a function of the layer thickness in $WS_2$ (c) and $WSe_2$ (f). Both show a nearly constant energy difference of ~ 0.4eV which corresponds to the splitting of the valence band edge.

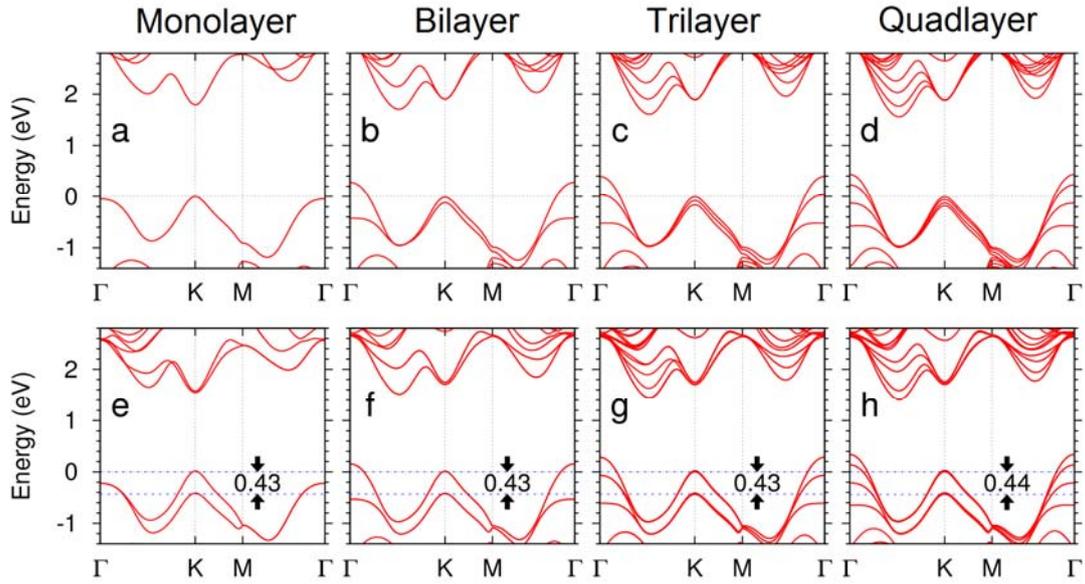

Figure 4. *Ab initio* calculated band structures for WS$_2$ ultrathin layers. (a)~(d): without spin-orbit coupling; (e)~(h): with spin-orbit coupling. The valence band splittings at K point are nearly constants, around 0.43eV for mono-, bi-, tri, and quad-layers.

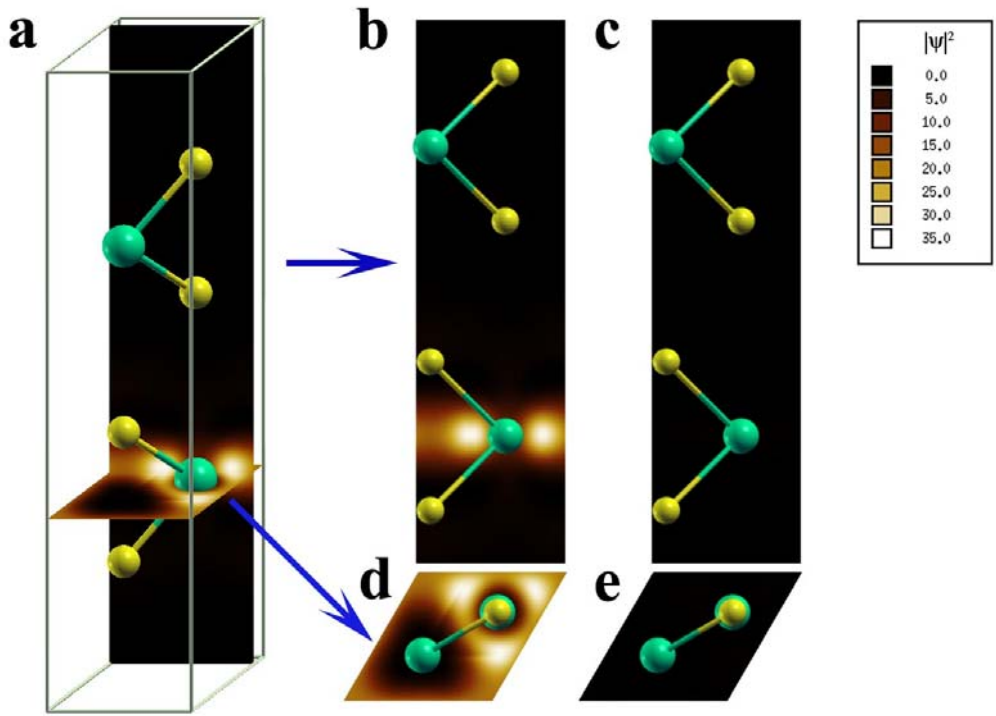

Figure 5. *Ab initio* calculated electron density maps for one valence band Bloch state at K point in WS$_2$ bilayer (indicated by the dashed arrow in Fig. 4f). (a) (b),(d) show the density of the spin down component of the wavefunction and (c),(e) show the density of the spin up component. The wavefunction is almost fully localized in the bottom layer and spin polarized in the down state. The other degenerate state at the same K point can be obtained by a spatial inversion plus a time reversal operation.